\documentclass[11pt]{article}
\usepackage[margin=1in]{geometry}
\usepackage{amsmath,amssymb}
\usepackage{graphicx}
\usepackage{booktabs}
\usepackage{longtable}
\usepackage{caption}
\usepackage[colorlinks=true,citecolor=blue,linkcolor=blue]{hyperref}
\usepackage{setspace}

\title{Beta Regression with Autoregressive Errors for Interrupted Time Series Analysis of Proportion and Rate Outcomes: A Simulation Study}

\author{Ariel Linden, DrPH\\[6pt]
\normalsize University of California, San Francisco\\
Department of Medicine\\
Division of Clinical Informatics \& Digital Transformation (DoC-IT)\\
San Francisco, CA, USA\\
ariel.linden@ucsf.edu}

\date{}

\begin{document}
\maketitle

\begin{abstract}
\noindent
Interrupted time series analyses (ITSA) of proportion and rate outcomes are frequently estimated using ordinary least squares regression, despite the bounded nature of these outcomes. When methods appropriate for bounded outcomes are used, the standard approach is a quasi-likelihood generalized linear model (GLM) with heteroskedasticity- and autocorrelation-consistent (HAC) standard errors. However, no existing estimator jointly models the beta-distributed conditional density and autoregressive (AR) error structure.
We introduce \texttt{betark}, a Stata implementation of a joint conditional maximum likelihood estimator for beta regression with AR(k) errors, based on a recursive substitution yielding a closed-form conditional beta likelihood with autoregressive dependence of arbitrary order. Unlike two-stage approaches, \texttt{betark} estimates the mean, precision, and AR($k$) coefficients jointly in a single likelihood, so reported standard errors already reflect the autocorrelation structure without separate correction. A Monte Carlo study compared \texttt{betark} against a quasi-binomial GLM with Newey--West HAC standard errors across AR(1)--AR(3) structures, three series lengths, and four effect sizes in a single-group ITSA design, evaluating bias, standard error ratio, 95\% coverage, Type I error, and power. Both methods were essentially unbiased. \texttt{betark} provided better-calibrated inference than the GLM+HAC comparator in most scenarios, with the largest advantages under highly persistent autocorrelation, where GLM+HAC Type I error exceeded 60\% at short series lengths. Misspecifying the AR order by one lag, and varying the starting mean and pre-intervention trend, had only modest effects on performance. However, \texttt{betark}'s own Type I error remained elevated under highly persistent AR(3) processes even at the largest series length examined.
\end{abstract}


\section{Introduction}

Interrupted time series analysis (ITSA) is widely used in health, social, and policy research to evaluate the effects of interventions, policy changes, and natural events on an outcome observed repeatedly over time \cite{campbell1966,shadish2002}. A single unit -- a hospital, a region, a health system -- is observed at regular intervals before and after an intervention, and the design's reliance on multiple pre- and post-intervention observations gives it strong internal validity relative to a simple before-after comparison, primarily through its control over regression to the mean
\cite{campbell1966,shadish2002,linden2007}, and generally has good
external validity, particularly when the unit of measurement is at the
population level or when results can be generalized to other units,
treatments, or settings \cite{shadish2002,linden2004}. Many outcomes evaluated in this design are proportions, rates, or percentages bounded within the unit interval. Throughout health services and policy research, common examples include readmission, infection, prescribing, and adherence rates, as well as other analogous population-level measures aggregated at each time point. 

When the outcome of interest is a proportion, fitting an ordinary least squares (OLS) model to the raw proportion—even when corrected for serial correlation—constitutes a form of model misspecification that is distinct from, but no less consequential than, ignoring autocorrelation. A linear model assumes a constant-variance, unbounded conditional mean function; for an outcome with support on $(0,1)$, this can produce fitted or predicted values outside the admissible range, and it disregards the well-documented relationship between the mean and variance of a bounded proportion, in which variance is largest near the center of the interval and shrinks toward the boundaries \cite{ferrari2004}. Beta regression directly models the conditional mean of a proportion outcome through a logit (or other) link function while simultaneously modeling its precision, avoiding both the boundary problem and the heteroskedasticity misspecification inherent to a linear model applied to bounded data \cite{ferrari2004}.

A second, independent form of model misspecification in ITSA arises from ignoring serial dependence. Time series data collected at regular intervals are routinely autocorrelated, and a substantial methodological literature -- largely developed for continuous, unbounded outcomes -- has documented that failing to account for autoregressive (AR) error structure produces biased standard errors, inflated Type I error, and unreliable confidence interval coverage in ITSA, with the severity of the problem increasing with the order and persistence of the underlying autocorrelation \cite{linden2026a,linden2026b,turner2021,bottomley2023}. The two methodological gaps—bounded support and autocorrelated errors—cannot be addressed independently. An investigator who corrects for serial correlation using methods developed for unbounded outcomes still ignores the mean-variance relationship inherent to bounded outcomes, while an investigator who fits a beta regression without accounting for serial dependence remains vulnerable to biased inference. No existing estimator jointly addresses both problems. 

The recursive substitution of Rocha and Cribari-Neto \cite{rocha2009}, extended to a partially linear setting by Ferreira, Figueroa-Z\'{u}\~{n}iga, and de Castro \cite{ferreira2015}, provides a closed-form conditional beta likelihood for an autoregressive process of arbitrary order $k$, in principle solving both problems jointly: the conditional mean is modeled on the logit scale, respecting the bounded support of the outcome, while the autoregressive structure is built directly into the same likelihood rather than corrected for after the fact. We introduce \texttt{betark}, a Stata implementation of this model estimated by joint conditional maximum likelihood, designed for time series analysis of proportion and rate outcomes with autoregressive errors of order $k$ \cite{lindenbetark}. To our knowledge, this recursive beta-AR($k$) formulation has not previously been implemented as a general-purpose regression command, nor evaluated for its finite-sample inferential properties -- bias, standard error calibration, confidence interval coverage, Type I error, and power -- under conditions representative of applied interrupted time series analysis. 

This study makes two contributions. First, we provide the first systematic Monte Carlo evaluation of the beta-AR($k$) model across autoregressive orders, autocorrelation structures, series lengths, and treatment effects representative of interrupted time series analysis. Second, we compare \texttt{betark} against the only readily available competing estimator for this combination of bounded outcome and autocorrelated errors: a quasi-binomial (fractional) generalized linear model with a logit link, paired with Newey--West heteroskedasticity- and autocorrelation-consistent (HAC) standard errors \cite{papke1996,newey1987,newey1994}, implemented in Stata as \texttt{glm} with the \texttt{vce(hac nwest \#)} option. Unlike \texttt{betark}, this comparator does not jointly model the bounded conditional mean and the autoregressive structure; it instead applies a post-estimation, nonparametric variance correction to a quasi-likelihood mean model that itself disregards the precision structure of the beta distribution. Whether this fractional-regression-plus-HAC approach provides adequate inference relative to a model that jointly addresses both sources of misspecification is the central empirical question motivating this study.

We compare \texttt{betark} with the fractional-regression-plus-HAC approach using Monte Carlo simulation spanning multiple autoregressive orders, autocorrelation structures, series lengths, and treatment effect sizes. Performance is evaluated using bias, standard error calibration, confidence interval coverage, Type I error, and statistical power. Supplementary analyses examine the consequences of AR-order misspecification and the sensitivity of the results to the starting mean and pre-intervention trend.

\section{Methods}

\subsection{The Beta Regression Model}

Let $y_t \in (0,1)$, $t = 1,\dots,T$, denote a continuous proportion or rate outcome observed sequentially over time. Following Ferrari and Cribari-Neto~\cite{ferrari2004}, we assume that, conditional on the information set $\mathcal{F}_{t-1}$, $y_t$ follows a beta distribution parameterized in terms of its conditional mean $\mu_t \in (0,1)$ and a precision parameter $\phi_t > 0$:
\begin{equation}
f(y_t \mid \mathcal{F}_{t-1}) = \frac{\Gamma(\phi_t)}{\Gamma(\mu_t \phi_t)\,\Gamma((1-\mu_t)\phi_t)}\, y_t^{\,\mu_t \phi_t - 1} (1-y_t)^{(1-\mu_t)\phi_t - 1}, \qquad 0 < y_t < 1,
\label{eq:betadensity}
\end{equation}
with $E(y_t \mid \mathcal{F}_{t-1}) = \mu_t$ and $\mathrm{Var}(y_t \mid \mathcal{F}_{t-1}) = \mu_t(1-\mu_t)/(1+\phi_t)$. The conditional mean is linked to a set of covariates through a strictly monotonic link function $g(\cdot): (0,1) \to \mathbb{R}$, typically the logit link:
\begin{equation}
g(\mu_t) = \mathrm{logit}(\mu_t) = \log\!\left(\frac{\mu_t}{1-\mu_t}\right) = x_t'\beta,
\label{eq:meaneq}
\end{equation}
where $x_t$ is a $q \times 1$ vector of covariates (including a constant) and $\beta$ is the corresponding coefficient vector. The precision parameter may likewise be modeled as constant, or as a function of covariates $z_t$ through a second link function, typically the log link:
\begin{equation}
\log(\phi_t) = z_t'\delta.
\label{eq:scaleeq}
\end{equation}
When $\phi_t = \phi$ is constant and $z_t$ contains only a constant, this reduces to the static beta regression model of Ferrari and Cribari-Neto~\cite{ferrari2004}, implemented in Stata as \texttt{betareg}. This static specification assumes that, conditional on $x_t$, observations are independent across $t$ -- an assumption frequently violated in time series data, where serially adjacent observations of a rate or proportion are typically correlated.

\subsection{Beta Regression with AR($k$) Errors}

To accommodate serial dependence, we follow the recursive substitution approach of Rocha and Cribari-Neto~\cite{rocha2009}, subsequently extended to a partially linear setting by Ferreira, Figueroa-Z\'{u}\~{n}iga, and de Castro~\cite{ferreira2015}. Define the disturbance $\xi_t = g(\mu_t) - x_t'\beta$, and let $\{\xi_t\}$ follow an autoregressive process of order $k$:
\begin{equation}
\xi_t = \sum_{i=1}^{k} \rho_i\left[g(y_{t-i}) - x_{t-i}'\beta\right] + r_t,
\label{eq:arrecursion}
\end{equation}
where $r_t$ is a martingale difference with $E(r_t \mid \mathcal{F}_{t-1}) = 0$. Substituting equation~\eqref{eq:arrecursion} into the mean equation~\eqref{eq:meaneq} yields the beta-AR($k$) conditional mean model:
\begin{equation}
g(\mu_t) = x_t'\beta + \sum_{i=1}^{k} \rho_i\left[g(y_{t-i}) - x_{t-i}'\beta\right], \qquad t = k+1,\dots,T.
\label{eq:betaark}
\end{equation}
Equation~\eqref{eq:betaark} differs from an additive autoregressive error structure of the kind used in Prais--Winsten or Newey--West corrected linear regression: the autoregressive feedback term is constructed from the \emph{realized}, link-transformed lagged outcomes $g(y_{t-i})$ rather than from a latent, mean-zero error series. This is a necessary consequence of the bounded support of $y_t$; an additive error term defined directly on the proportion scale could push $\mu_t$ outside $(0,1)$.

The joint conditional log-likelihood, given the first $k$ observations, is
\begin{equation}
\ell(\theta) = \sum_{t=k+1}^{T} \log f(y_t \mid \mathcal{F}_{t-1}; \theta),
\label{eq:condll}
\end{equation}
where $\theta = (\beta', \delta', \rho')'$, $\rho = (\rho_1,\dots,\rho_k)'$, and $f(\cdot)$ is the beta density of equation~\eqref{eq:betadensity} evaluated at $\mu_t$ and $\phi_t$ as implied by equations~\eqref{eq:scaleeq} and~\eqref{eq:betaark}. Unlike two-stage approaches that estimate $\beta$ first (e.g., by ordinary least squares or static beta regression) and the autoregressive parameters $\rho$ in a second stage conditional on the first-stage residuals -- the approach used, for example, by INAR-based count time series models -- equation~\eqref{eq:condll} is maximized jointly over $\beta$, $\delta$, and $\rho$ simultaneously. Consequently, the resulting covariance matrix of $\hat\theta$ is the ordinary asymptotic covariance of a joint maximum likelihood estimator and already reflects the estimated autocorrelation structure; no separate sandwich or two-stage correction to the standard errors of $\hat\beta$ is required or applied.

\subsubsection{Estimation}

The model is implemented in the Stata package \texttt{betark}. Estimation proceeds by maximizing equation~\eqref{eq:condll} using Stata's \texttt{moptimize()} optimization suite, the same optimization engine used by official estimation commands, including \texttt{betareg}. The mean equation, scale equation, and each autoregressive coefficient $\rho_1,\dots,\rho_k$ are specified as separate equations within \texttt{moptimize()}'s \texttt{"lf"} evaluator, which returns the per-observation log-likelihood contributions defined by equation~\eqref{eq:condll}; observations $t \le k$ within each contiguous time segment contribute zero to the likelihood, since the AR($k$) recursion requires $k$ realized lags before it is defined. For panel or multiple-segment data, the AR($k$) recursion restarts at the first $k$ observations following any gap in the time variable or change in panel identifier, mirroring the segment-handling logic used in \texttt{praisk}, the companion time-series package for continuous outcomes \cite{linden2026b}. Starting values for $\beta$ and $\delta$ are obtained from a static \texttt{betareg} fit ($\rho = 0$); $\rho$ is initialized at $0.1/k$ for each lag.

\subsection{Comparator: Quasi-Binomial GLM with Newey--West HAC Standard Errors}

As a comparator representing the principal estimation approach currently available for bounded outcomes with autocorrelated errors, we evaluate a generalized linear model with a binomial family and logit link, fit by quasi-maximum likelihood -- the standard ``fractional regression'' approach for continuous proportion outcomes bounded on $(0,1)$ \cite{papke1996} -- combined with a Newey--West heteroskedasticity- and autocorrelation-consistent (HAC) variance correction \cite{newey1987,newey1994}. This comparator is analogous to the ordinary-least-squares-with-Newey--West (OLS-NW) estimator evaluated in prior simulation work on interrupted time series analysis with continuous outcomes \cite{linden2026a,linden2026b}, adapted here to the logit-linked quasi-likelihood appropriate for a bounded outcome.

For the linear predictor $\eta_t = x_t'\gamma$ with $g(\mu_t) = \eta_t$, the quasi-likelihood estimator $\hat\gamma$ solves the binomial score equations
\begin{equation}
\sum_{t=1}^{T} x_t \left(y_t - \mu_t(\eta_t)\right) = 0,
\label{eq:glmscore}
\end{equation}
which yields a consistent estimator of $\gamma$ for any conditional mean model of the form $g(\mu_t) = x_t'\gamma$, regardless of the true conditional distribution of $y_t$, provided the mean is correctly specified \cite{papke1996}. Because this quasi-likelihood approach makes no assumption about the form of serial dependence in $y_t$, inference is instead based on the Newey--West HAC covariance estimator. Let $\hat{u}_t = y_t - \hat\mu_t$ denote the working residual. The HAC long-run covariance matrix is estimated as
\begin{equation}
\hat{S}_{NW} = \hat\Gamma(0) + \sum_{j=1}^{L} w_j \left[\hat\Gamma(j) + \hat\Gamma(j)'\right],
\label{eq:hac}
\end{equation}
where $\hat\Gamma(j) = T^{-1}\sum_{t=j+1}^{T} \hat{u}_t \hat{u}_{t-j}\, x_t x_{t-j}'$ and $w_j = 1 - j/(L+1)$ are Bartlett kernel weights \cite{newey1987}. The bandwidth $L$ is matched to the true AR order in the corresponding simulation cell -- i.e., \texttt{vce(hac nwest \#)} with \texttt{\#} set equal to $k$ -- representing the idealized case in which the analyst correctly identifies the order of serial dependence, even though the HAC correction itself does not require this. The full HAC covariance matrix of $\hat\gamma$ is
\begin{equation}
\widehat{\mathrm{Var}}_{NW}(\hat\gamma) = \left(X'WX\right)^{-1} X'\hat{S}_{NW}X \left(X'WX\right)^{-1},
\label{eq:hacvar}
\end{equation}
where $W$ is the diagonal weight matrix from the binomial iteratively reweighted least squares fit. This comparator is implemented in Stata via \texttt{glm \dots, family(binomial) link(logit) vce(hac nwest \#)}.

\subsection{The Single-Group Interrupted Time Series Model}

When there is only one group under study (no comparison group) and only a single intervention period, the standard interrupted time series analysis (ITSA) regression model assumes the following form \cite{linden2015}:
\begin{equation}
Y_t = \beta_0 + \beta_1 T_t + \beta_2 X_t + \beta_3 X_t T_t + \epsilon_t,
\label{eq:itsa}
\end{equation}
where $Y_t$ is the outcome variable measured at each equally spaced time point $t$; $T_t$ is the time since the start of the study; $X_t$ is a dummy (indicator) variable representing the intervention (0 in pre-intervention periods, 1 otherwise); and $X_t T_t$ is an interaction term. $\beta_0$ represents the intercept, or starting level, of the outcome variable. $\beta_1$ is the slope, or trend, of the outcome variable prior to the intervention. $\beta_2$ represents the change in the level of the outcome occurring in the period immediately following the intervention. $\beta_3$ represents the difference between the pre-intervention and post-intervention slopes of the outcome. A significant $\beta_2$ indicates an immediate treatment effect, and a significant $\beta_3$ indicates a treatment effect that accumulates over time \cite{linden2015}.

Equation~\eqref{eq:itsa} is the linear-model analogue of the beta-AR($k$) mean equation, equation~\eqref{eq:dgplinpred}, of the present study: $\beta_0$ corresponds to $\eta_0$, $\beta_1$ to $b_{\mathrm{pre}}$, $\beta_2$ to $b_{\mathrm{step}}$, and $\beta_3$ to $b_{\mathrm{post}}-b_{\mathrm{pre}}$, the slope-change coefficient. Consistent with this correspondence, $\beta_3$ -- the difference between pre- and post-intervention trend -- is the treatment effect parameter evaluated throughout the simulation study reported below.

\subsection{Simulation Strategy}

\subsubsection{Data-generating process}

Artificial series were generated under the same beta-AR($k$) model described in equations~\eqref{eq:betadensity}--\eqref{eq:betaark}, using a single-group interrupted time series design. For a series of length $T$ with intervention initiated at the midpoint, $\tau = \lfloor T/2 \rfloor$, the linear predictor was specified as
\begin{equation}
\eta_t = \eta_0 + b_{\mathrm{pre}}\, t + b_{\mathrm{step}}\, D_t + (b_{\mathrm{post}} - b_{\mathrm{pre}})\, D_t (t-\tau), \qquad D_t = \mathbb{1}(t \ge \tau),
\label{eq:dgplinpred}
\end{equation}
where $\eta_0 = \mathrm{logit}(\mu_0)$, $\mu_0$ is the starting mean, $b_{\mathrm{pre}}$ and $b_{\mathrm{post}}$ are the logit-scale pre- and post-intervention slopes, and $b_{\mathrm{step}}$ is a logit-scale immediate level-change coefficient (set to zero throughout the present design). The series was then generated sequentially according to the AR($k$) recursion of equation~\eqref{eq:betaark}: for $t=1,\dots,k$, $\mu_t = g^{-1}(\eta_t)$ (no autoregressive adjustment is defined for the first $k$ observations, which serve to initialize the recursion); for $t=k+1,\dots,T$,
\begin{equation}
\mu_t = g^{-1}\!\left(\eta_t + \sum_{i=1}^{k} \rho_i \left[g(y_{t-i}) - \eta_{t-i}\right]\right), \qquad y_t \sim \mathrm{Beta}(\mu_t \phi, (1-\mu_t)\phi),
\label{eq:dgprecursion}
\end{equation}
drawn sequentially so that each $y_t$ depends on the realized values of $y_{t-1},\dots,y_{t-k}$, consistent with the recursive substitution of Rocha and Cribari-Neto~\cite{rocha2009} and Ferreira et al.~\cite{ferreira2015} on which the \texttt{betark} likelihood is based.

The treatment effect of interest is the slope-change parameter, $b_{\mathrm{post}} - b_{\mathrm{pre}}$, i.e., the coefficient on the interaction $D_t(t-\tau)$ in equation~\eqref{eq:dgplinpred}, paralleling the difference-in-differences-in-trend parameterization used in the related Prais--Winsten and Newey--West simulation studies for continuous and panel time series outcomes \cite{linden2026a,linden2026b,linden2026c}. Because percentage effect sizes are more readily interpretable than logit-scale coefficients, $b_{\mathrm{pre}}$, $b_{\mathrm{step}}$, and $b_{\mathrm{post}}$ were not specified directly; instead, each was solved so that the corresponding total percentage change in $\mu_t$, expressed relative to $\mu_0$, was realized exactly at the end of its respective segment. Specifically, given a target pre-intervention percentage change $\delta_{\mathrm{pre}}$,
\begin{equation}
b_{\mathrm{pre}} = \frac{\mathrm{logit}\!\left(\mu_0(1+\delta_{\mathrm{pre}})\right) - \eta_0}{\tau},
\label{eq:bpre}
\end{equation}
so that the cumulative pre-intervention drift from $t=0$ to $t=\tau$ corresponds to a total change of $100\,\delta_{\mathrm{pre}}\%$ of $\mu_0$. The immediate level-change coefficient was solved analogously as a one-period shift relative to the counterfactual pre-trend value at $\tau$, $\mu_\tau^{\mathrm{cf}} = g^{-1}(\eta_0 + b_{\mathrm{pre}}\tau)$:
\begin{equation}
b_{\mathrm{step}} = \mathrm{logit}\!\left(\mu_\tau^{\mathrm{cf}} + \mu_0 \delta_{\mathrm{step}}\right) - \mathrm{logit}\!\left(\mu_\tau^{\mathrm{cf}}\right),
\label{eq:bstep}
\end{equation}
and the post-intervention slope was solved so that the cumulative post-intervention drift, from $\mu_\tau = g^{-1}(\eta_0 + b_{\mathrm{pre}}\tau + b_{\mathrm{step}})$ to the end of the series, corresponds to a total change of $100\,\delta_{\mathrm{post}}\%$ of $\mu_0$ over the remaining $T-1-\tau$ periods:
\begin{equation}
b_{\mathrm{post}} = \frac{\mathrm{logit}\!\left(\mu_\tau + \mu_0 \delta_{\mathrm{post}}\right) - \mathrm{logit}(\mu_\tau)}{T-1-\tau}.
\label{eq:bpost}
\end{equation}
This anchoring -- a total cumulative percentage change over each segment, rather than a constant per-period percentage change -- was adopted because the logit link compresses near the boundaries of $(0,1)$. A constant per-period logit-scale slope, if calibrated to a fixed percentage change at $t=1$, compounds approximately geometrically and can drive $\mu_t$ to the boundary well before the end of a long segment, whereas anchoring to the total change at the segment's endpoint yields a smoothly varying trajectory of the intended magnitude throughout. All three coefficients are checked at the point of construction to confirm that every implied mean remains strictly within $(0,1)$.

The dispersion of the simulated series was parameterized through a target coefficient of variation of $y_t$ at the starting mean, $\mathrm{cv} = \mathrm{SD}(y_t)/\mu_0$, rather than through the beta precision parameter $\phi$ directly, since a fixed $\phi$ implies markedly different relative dispersion depending on where $\mu_0$ lies between the boundaries of $(0,1)$ (the variance $\mu(1-\mu)/(1+\phi)$ is maximized at $\mu=0.5$ and shrinks toward either boundary). Equating $\mathrm{Var}(y_t) = \mu_0(1-\mu_0)/(1+\phi)$ with $(\mathrm{cv}\cdot\mu_0)^2$ and solving for $\phi$ gives
\begin{equation}
\phi = \frac{1-\mu_0}{\mu_0\, \mathrm{cv}^2} - 1.
\label{eq:cvtophi}
\end{equation}
A value of $\mathrm{cv} = .05$ was used throughout the primary design, corresponding to $\phi \approx 3{,}599$ at $\mu_0 = .10$ via equation~\eqref{eq:cvtophi}; this value was chosen based on a preliminary Monte Carlo evaluation of the recursive beta-AR($k$) generating process's small-sample bias as a function of dispersion (Appendix~A), which confirmed negligible bias in the step-effect coefficient at this and lower noise levels.

The starting mean (intercept) $\mu_0$ was fixed at $.10$ across the primary design, reflecting the low base rates typical of the rate and proportion outcomes (e.g., readmission rates, infection rates) that motivate this work, and representing a more demanding test case for both estimators than a centrally located mean, since the precision of the beta distribution -- and consequently both the small-sample behavior of \texttt{betark}'s recursive likelihood and the variance of the GLM working residuals -- depends on the distance of $\mu_t$ from the boundary of $(0,1)$. The pre-intervention trend ($\delta_{\mathrm{pre}}$) was held at zero throughout the primary design to isolate the post-intervention trend effect; the post-intervention trend ($\delta_{\mathrm{post}}$) was varied as described below. No immediate level change was specified ($\delta_{\mathrm{step}} = 0$), consistent with an intervention expected to affect the outcome through a gradual change in trend rather than an abrupt shift in level.

\subsubsection{Autoregressive order and scenarios}

For each AR order $k \in \{1,2,3\}$, three autocorrelation scenarios were specified to represent qualitatively distinct patterns of serial dependence, following the scenario design of Linden~\cite{linden2026b,linden2026c}:
\begin{itemize}
\item \textbf{Scenario 1 (mild positive):} AR(1) $\rho = 0.4$; AR(2) $\rho = (0.4, 0.2)$; AR(3) $\rho = (0.4, 0.2, 0.1)$.
\item \textbf{Scenario 2 (oscillatory):} AR(1) $\rho = -0.4$; AR(2) $\rho = (0.5, -0.4)$; AR(3) $\rho = (0.7, -0.3, 0.15)$.
\item \textbf{Scenario 3 (high persistent):} AR(1) $\rho = 0.7$; AR(2) $\rho = (0.7, 0.2)$; AR(3) $\rho = (0.6, 0.25, 0.1)$.
\end{itemize}
All nine parameter combinations satisfy the stationarity condition that all eigenvalues of the corresponding AR companion matrix lie strictly inside the unit circle, confirmed numerically prior to simulation. The negative second-order coefficients in Scenario 2 produce complex companion-matrix roots, yielding genuinely oscillatory (cyclical) autocovariance functions rather than purely monotonic decay; Scenario 3's coefficients approach, but remain within, the stationarity boundary, producing near-unit-root persistence.

\subsubsection{Series length and effect size}

Three series lengths were examined: $T \in \{100, 200, 400\}$. Four post-intervention trend effect sizes were examined, expressed as a percentage of the starting mean and realized as a total cumulative change in $\mu_t$ over the post-intervention period: $0\%$ (the null condition, for Type I error and coverage evaluation), $25\%$ (small), $50\%$ (medium), and $100\%$ (large).

\subsubsection{Estimation methods}

For each combination of $T$, autocorrelation scenario, and effect size, two estimators were fit to the same generated dataset: \texttt{betark} with \texttt{lag()} set to the true AR order $k$, and the quasi-binomial GLM with \texttt{vce(hac nwest \#)}, \texttt{\#} likewise set to $k$. Both methods were therefore correctly specified with respect to the true autoregressive order in the primary analysis; misspecification is examined separately (Section~\ref{sec:misspec}).

\subsubsection{Replications and performance measures}

Each of the $3 \times 9 \times 4 = 108$ design cells was replicated $2000$ times. Following Burton et al.~\cite{burton2006}, five performance measures were computed for the slope-change coefficient: statistical power (the rejection rate under each non-null effect size), 95\% confidence interval coverage, Type I error rate (the rejection rate at $H_0: \beta_{\mathrm{post}}-\beta_{\mathrm{pre}} = 0$ under the $0\%$-effect null condition), percentage bias, and the standard error ratio (mean model-based standard error divided by the empirical standard deviation of the point estimates across replications). All hypothesis tests used two-sided Wald tests based on each estimator's own reported standard error. Bias, the standard error ratio, and confidence interval coverage are reported only for the non-null effect sizes ($25\%$, $50\%$, $100\%$), averaged across these three; the null-effect ($0\%$) replications are reserved exclusively for Type I error evaluation and are not pooled with the non-null replications in any other performance measure, since doing so would conflate a measure of estimation/calibration quality with a measure of false-positive behavior. Table~\ref{tab:design} summarizes the full simulation design.

\subsubsection{Misspecification analysis}
\label{sec:misspec}

A supplementary simulation examined the consequences of AR order misspecification for \texttt{betark}. For each true AR order $k \in \{1,2,3\}$ under the mild-positive autocorrelation scenario, data were generated under the correct order and fit with three lag specifications: the true order, one lag under-specified, and one lag over-specified (where applicable). This design was crossed with all three series lengths, holding the post-intervention trend effect at $50\%$. The GLM+HAC comparator was not included in this analysis, since its HAC correction does not require specifying an AR order.

\subsubsection{Sensitivity analysis}

A second supplementary simulation examined whether the primary findings are sensitive to the choice of starting mean and the presence of a non-zero pre-intervention trend. Holding the autoregressive structure fixed at AR(1) Scenario 1 ($\rho = 0.4$) and the series length fixed at $T=200$, the starting mean was varied over $\mu_0 \in \{.05, .10, .30, .50\}$ and the pre-intervention trend was varied over $\delta_{\mathrm{pre}} \in \{-25\%, 0\%, 25\%\}$, crossed with two post-intervention trend effect sizes ($0\%$ and $50\%$). Both \texttt{betark} and the GLM+HAC comparator were evaluated under each combination.

\section{Results}

\subsection{Statistical Power}
\label{sec:power}

Statistical power at the medium ($50\%$) and large ($100\%$) effect sizes was at or near $1.0$ for both methods in the majority of scenarios, providing limited discriminating information (Tables~\ref{tab:power50}--\ref{tab:power100}). The small-effect-size ($25\%$) results (Table~\ref{tab:power25}) were more informative. In the two high-persistence scenarios, the GLM+HAC comparator showed higher raw power than \texttt{betark} (e.g., $0.466$ versus $0.720$ for AR(2) high-persistent autocorrelation at $T=100$). This apparent advantage must be interpreted in light of the Type I error results in Section~\ref{sec:typeI}: because a Wald test rejects more often whenever its standard error is too small, regardless of whether the null hypothesis is true, the GLM+HAC comparator's elevated power in these cells is mechanically linked to its elevated Type I error in the same cells, rather than reflecting a genuine sensitivity advantage. We further note a non-monotonic pattern in \texttt{betark}'s small-effect power for the two high-persistence scenarios: power for AR(2) and AR(3) high-persistent autocorrelation did not increase monotonically with $T$ (e.g., $0.542 \to 0.400 \to 0.336$ for AR(3) high persistent, $T = 100, 200, 400$). This pattern is consistent with -- and a direct consequence of -- the simultaneous improvement in Type I error documented in Section~\ref{sec:typeI}: as \texttt{betark}'s standard errors become better calibrated with increasing $T$ in these difficult scenarios, the inflated rejection rate that had been partially attributable to anticonservative inference recedes, reducing power for genuinely small effects until the sample size is large enough for the gain in genuine sensitivity to dominate.

\subsection{95\% Confidence Interval Coverage}
\label{sec:coverage}

\texttt{betark} achieved higher 95\% confidence interval coverage than the GLM+HAC comparator in seven of nine autocorrelation scenarios (Table~\ref{tab:coverage}). The advantage was largest under high-persistent autocorrelation: at $T=100$, coverage for AR(2) high-persistent autocorrelation was 0.677 for \texttt{betark} versus 0.429 for the GLM+HAC comparator; for AR(3) high-persistent autocorrelation, 0.565 versus 0.345. \texttt{betark}'s coverage advantage in these scenarios widened, or at minimum did not narrow, with increasing $T$, whereas the GLM+HAC comparator's coverage in the same scenarios improved only modestly with series length. In the two oscillatory scenarios, the GLM+HAC comparator showed higher nominal coverage than \texttt{betark} (e.g., 0.975 versus 0.948 for AR(1) oscillatory autocorrelation at T = 100); however, this reflects conservative (wider) standard errors rather than superior calibration (Section~\ref{sec:seratio}).

\subsection{Type I Error}
\label{sec:typeI}

Type I error rates, evaluated under the true null ($0\%$ post-intervention trend effect), showed the same pattern as coverage, as expected given the direct relationship between the two measures (Table~\ref{tab:typeI}). \texttt{betark}'s Type I error exceeded the nominal $5\%$ level in most scenarios, with the degree of inflation increasing with AR order and autocorrelation persistence; however, the GLM+HAC comparator's inflation was consistently larger in the corresponding cells. Under AR(2) high-persistent autocorrelation at $T=100$, Type I error was $30.8\%$ for \texttt{betark} versus $60.2\%$ for the GLM+HAC comparator; under AR(3) high-persistent autocorrelation, $43.0\%$ versus $65.4\%$. \texttt{betark}'s Type I error declined with increasing $T$ in every scenario examined (e.g., from $43.0\%$ at $T=100$ to $23.0\%$ at $T=400$ for AR(3) high-persistent autocorrelation), consistent with asymptotic convergence toward nominal coverage. The GLM+HAC comparator's Type I error in the corresponding high-persistence cells showed little or no improvement with $T$, and in one scenario (AR(3) high persistent) increased modestly as $T$ grew from 100 to 400.

\subsection{Bias}
\label{sec:bias}

Across all nine autocorrelation scenarios, three series lengths, and three non-null trend effect sizes (null-effect replications are reserved for Type I error; Section~\ref{sec:typeI}), both \texttt{betark} and the GLM+HAC comparator were essentially unbiased (Table~\ref{tab:bias}). Bias for \texttt{betark} ranged from $-0.0001$ to $0.0004$ across all 27 (scenario $\times$ $T$) cells, averaged across non-null effect sizes; bias for the GLM+HAC comparator was of comparable magnitude ($-0.0001$ to $0.0006$). Neither method showed a systematic directional bias, nor any tendency for bias to increase with AR order, autocorrelation persistence, or series length. This confirms that any differences in inferential performance between the two methods (Sections~\ref{sec:power}--\ref{sec:typeI}) arise from differences in variance estimation rather than from systematic error in point estimation, consistent with the analogous finding for Prais--Winsten and Newey--West estimators applied to continuous outcomes \cite{linden2026a,linden2026b,linden2026c}.

\subsection{Standard Error Ratio}
\label{sec:seratio}

The standard error ratio (Table 8), likewise averaged across non-null trend effect sizes only, explains the coverage and Type I error patterns reported above. \texttt{betark}'s standard error ratio was closer to the ideal value of 1.0 than the GLM+HAC comparator's in seven of nine autocorrelation scenarios. The two exceptions were both oscillatory (negative-$\rho$) scenarios, in which the GLM+HAC comparator's ratio exceeded 1.0 (e.g., 1.14--1.19 under AR(1) oscillatory autocorrelation), indicating conservative (overly large) standard errors, while \texttt{betark}'s ratio remained close to 1.0 throughout. In the remaining seven scenarios -- all mild-positive and high-persistent conditions -- the GLM+HAC comparator's ratio was substantially below 1.0, indicating anticonservative (too small) standard errors, most severely under high-persistent autocorrelation (e.g., 0.22--0.23 across series lengths for AR(3) high-persistent autocorrelation, versus 0.55--0.77 for \texttt{betark} under the same conditions). Notably, the GLM+HAC comparator's standard error ratio in the high-persistence scenarios did not improve materially with increasing $T$, whereas \texttt{betark}'s generally did, though not monotonically in every cell.

\subsection{Misspecification Analysis}

The misspecification analysis examined the consequences of fitting \texttt{betark} with an AR order one lag below or one lag above the true generating order, under the mild-positive autocorrelation scenario (Appendix~B). At $T=100$, coverage under correct AR(3) specification was $0.782$; under-specification (fitting AR(2)) and over-specification (fitting AR(4)) produced coverage of $0.772$ and $0.768$, respectively -- differences of one to two percentage points, small relative to the 20-plus percentage point differences associated with autocorrelation persistence and series length documented in Section~\ref{sec:coverage}. This pattern held across all three true AR orders and all three series lengths examined: misspecifying the AR order by a single lag carried a substantially smaller inferential cost than the choice of series length or the underlying persistence of the autocorrelation process itself.

\subsection{Sensitivity Analysis}

The sensitivity analysis examined whether the primary findings, derived under a fixed starting mean of $\mu_0 = .10$ and a flat pre-intervention trend, generalize to other starting means and to a sloped pre-intervention baseline (Appendix~C). Holding the autoregressive structure fixed at AR(1) Scenario 1 and $T=200$, \texttt{betark}'s coverage remained consistently in the range of $0.91$--$0.95$ across all four starting means ($\mu_0 \in \{.05,.10,.30,.50\}$) and all three pre-intervention trend conditions ($-25\%, 0\%, 25\%$), with no evidence that moving the starting mean away from the boundary, or introducing a sloped pre-intervention counterfactual, degraded calibration. The GLM+HAC comparator's coverage was consistently five to ten percentage points lower than \texttt{betark}'s across the same cells, reproducing the head-to-head finding of Section~\ref{sec:coverage} even in this comparatively favorable, low-persistence corner of the design space.

\section{Illustrative Example}

\subsection{Background and Study Design}

To illustrate the practical consequences of estimator choice under high-persistence autocorrelation, we analyze an artificial dataset reflecting a realistic disease management study. Disease management programs target individuals with chronic conditions or elevated risk through structured behavioral and clinical interventions \cite{linden2008,linden2003dmaa,kullgren2018}. Prediabetes is a common target given the effectiveness of lifestyle interventions in preventing progression to type 2 diabetes \cite{biuso2007}.

The artificial study involves a single primary care practice whose enrolled patients with prediabetes were fitted with continuous glucose monitors and tracked daily for 300 days. The outcome is the daily proportion of enrolled patients whose fasting glucose reading meets or exceeds the clinical target threshold of 100~mg/dL -- the upper bound of the normal fasting range and the threshold separating controlled from uncontrolled glycemic status -- aggregated across the practice's patient panel each day. Unlike the continuous mg/dL outcome examined in related work on the linear ITSA model \cite{linden2026b}, this outcome is itself a bounded proportion, directly motivating the beta-AR($k$) approach developed here. At day 150, the practice introduced a comprehensive lifestyle intervention program; patients were monitored for a further 150 days following the intervention.

Data were generated under the single-group ITSA model of equation~\eqref{eq:itsa} -- equivalently, the beta-AR($k$) data-generating process of equations~\eqref{eq:dgplinpred}--\eqref{eq:dgprecursion} -- using the parameterization described in Section~2.4. The starting proportion was set to $\mu_0 = .35$ (35\% of the patient panel above the glycemic threshold at baseline), within a clinically plausible range for an enrolled prediabetic population. The pre-intervention trend was set to a $10\%$ total increase in the elevated-glucose proportion over the 150-day baseline period, reflecting the gradual glycemic deterioration typical of untreated prediabetes. No immediate level change was specified at the intervention point ($\delta_{\mathrm{step}}=0\%$), consistent with a lifestyle intervention whose effects accumulate gradually rather than acting immediately. The post-intervention trend was set to a $15\%$ total \emph{decrease} in the elevated-glucose proportion over the 150-day follow-up period, reflecting a modest, clinically plausible improvement in glycemic control following the intervention -- deliberately chosen to be a smaller effect than would be trivially detected by either estimator regardless of standard error calibration, so that any difference in inferential conclusions between methods would more plausibly reflect a genuine consequence of standard error miscalibration rather than be obscured by an effect large enough to remain significant under either method's standard error. The true treatment effect -- $\beta_3$ in equation~\eqref{eq:itsa}, the slope-change coefficient $b_{\mathrm{post}}-b_{\mathrm{pre}}$ -- is therefore the logit-scale equivalent of a $10\%$ increase reversing into a $15\%$ decrease. Dispersion was set to $\mathrm{cv}=.05$, matching the primary simulation.

High persistent positive autocorrelation was specified to reflect plausible serial dependence in longitudinal, practice-level aggregate glycemic measurements -- the same condition under which the primary simulation results showed the largest divergence between estimators. Three datasets were generated using identical parameters and a common random seed, differing only in AR order: AR(1) ($\rho=0.7$), AR(2) ($\rho=(0.7,0.2)$), and AR(3) ($\rho=(0.6,0.25,0.1)$), matching the high-persistent scenario (Scenario 3) examined throughout the primary simulation. Each dataset was analyzed using both \texttt{betark} and the quasi-binomial GLM with Newey--West HAC standard errors, with the lag order in each case matched to the AR order of the generating process. Because the data-generating process is identical across estimators for a given AR order, the fitted trajectories do not differ between methods; the methods diverge only in their standard errors and inferential conclusions.

\subsection{Results}

\begin{figure}[htbp]
\centering
\includegraphics[width=\textwidth, height=0.6\textheight, keepaspectratio]{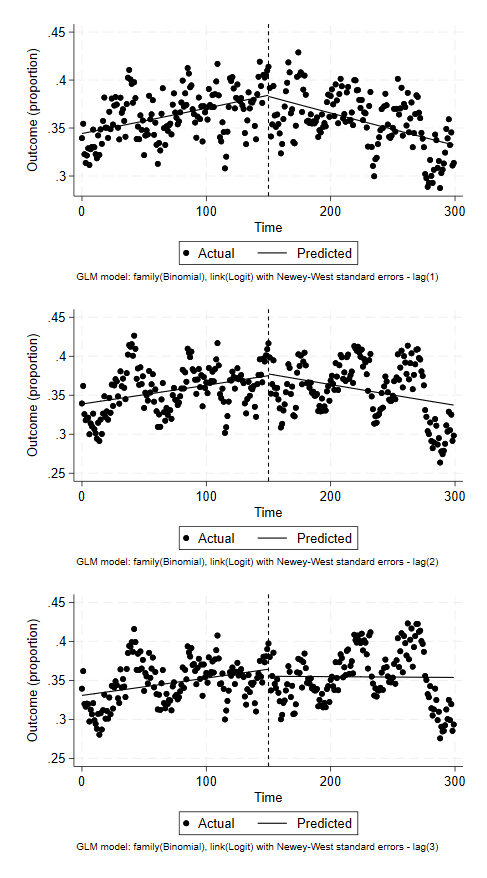}
\caption{Proportion of patients above the glycemic threshold (observed, gray circles) with \texttt{betark}-fitted trajectory (black line), by AR order. The dashed vertical line marks the intervention at day 150. Because the data-generating process is identical across estimators for a given AR order, the fitted trajectories shown here are from \texttt{glm}+HAC; \texttt{betark} fitted values are indistinguishable and are not shown separately.}
\label{fig:applied}
\end{figure}

Figure~\ref{fig:applied} displays the observed daily proportion series and the fitted trajectories for the AR(1) -- AR(3) orders using the GLM-HAC model (because the data-generating process is identical across estimators for a given AR order, the fitted values for \texttt{betark} are indistinguishable and are therefore not shown separately). The post-intervention decline in the proportion of patients above the glycemic threshold is visible across all three panels, consistent with the intervention's intended effect. Table~\ref{tab:applied} presents the slope-change coefficient estimates, standard errors, 95\% confidence intervals, and $p$-values for each AR order and estimation method. As in the primary simulation, these results represent a single realization; the corresponding Monte Carlo results (Section~3) confirm that both methods are approximately unbiased on average, so the patterns illustrated here in standard error calibration are the ones expected to generalize, not the point estimates themselves.

Under AR(1), the two methods agree closely: similar point estimates, both highly significant. As AR order increases, the two methods diverge -- not in the direction of one method crossing the conventional significance threshold while the other does not, but in the more fundamental sense that each method's expressed confidence in its own estimate becomes progressively less consistent with the other's. Under AR(2), \texttt{betark}'s point estimate ($-0.00276$) is larger in magnitude than the GLM+HAC comparator's ($-0.00222$), yet \texttt{betark}'s result is markedly less certain ($p=0.039$, barely below the conventional threshold) than the GLM+HAC comparator's ($p<0.001$) -- the opposite of what would be expected if the larger point estimate reflected a more clearly detectable effect. Under AR(3) this pattern sharpens into an outright inversion: the GLM+HAC comparator's point estimate ($-0.00103$) is now \emph{smaller} in magnitude than \texttt{betark}'s ($-0.00182$), and yet it reports the \emph{smaller} $p$-value ($0.134$ versus $0.283$). A weaker effect estimate appearing more statistically convincing than a stronger one is not a coincidence of this particular draw; it is the direct, mechanical consequence of the standard error miscalibration documented in the primary simulation (Section~\ref{sec:seratio}) -- the GLM+HAC comparator's standard error shrinks relative to its own sampling variability as AR order and persistence increase, inflating its apparent precision independently of the actual magnitude of the estimated effect. Neither method crosses the conventional significance threshold differently from the other in this particular realization, but the inversion itself is arguably a more concrete illustration of the paper's central finding than a simple significant/non-significant split would have been: an investigator relying on the GLM+HAC comparator's reported precision alone, without independent knowledge of the true autocorrelation structure, would have no way of recognizing that its apparent confidence is, in this case, the symptom of miscalibration rather than evidence of a well-estimated effect.

\section{Discussion}

The fundamental finding of this study is that \texttt{betark} provides better-calibrated inference than a quasi-binomial GLM with Newey--West HAC standard errors across the large majority of conditions examined, and that the size of this advantage is not uniform: it depends on the order and persistence of the autoregressive process in a way that has direct practical implications for applied interrupted time series analysis of proportion and rate outcomes. The two methods are not simply better or worse versions of each other. They address autocorrelated, bounded outcome data through fundamentally different mechanisms -- a joint conditional likelihood that models the autoregressive structure directly, against a nonparametric post-estimation variance correction applied to a model that disregards the beta distribution's precision structure entirely -- and the consequences of that difference widen precisely as the inferential stakes increase.

\subsection{Statistical Power}

The GLM+HAC comparator's higher raw power in the two high-persistence scenarios is not a sign of statistical efficiency. It is a direct consequence of the same standard error underestimation responsible for its inflated Type I error in the identical cells: a Wald test rejects more often whenever its standard error is too small, regardless of whether the null hypothesis is true. A researcher who selects the GLM+HAC comparator for its apparent power advantage under persistent autocorrelation is, in practice, accepting an elevated false-positive rate as the price, exactly as has been documented for the analogous OLS-Newey--West estimator in continuous-outcome interrupted time series designs \cite{linden2026b}. A fair, size-adjusted comparison of statistical power would require recalibrating each method's rejection threshold to its own empirical null distribution. Because this was not undertaken here, the observed power differences should not be interpreted as a formal comparison of statistical efficiency. The non-monotonic small-effect power observed for \texttt{betark} in the two high-persistence scenarios (Section~\ref{sec:power}) is, on inspection, a reassuring pattern rather than a  concerning one: as \texttt{betark}'s standard errors become better calibrated with increasing $T$ in these difficult scenarios, the portion of its rejection rate that had been attributable to anticonservative inference recedes before the portion attributable to genuine sensitivity has fully caught up. This is the opposite of what would be expected if \texttt{betark}'s power were itself an artifact of miscalibration: a genuinely anticonservative estimator's power and its Type I error would decline together, not show the orderly trade-off observed here.

\subsection{Coverage and Type I Error}

The coverage and Type I error results indicate that the two estimators degrade in qualitatively similar, but quantitatively very different, ways as autocorrelation persistence and AR order increase. Under mild positive and oscillatory autocorrelation, both methods showed only modest departures from nominal calibration, and in the oscillatory scenarios the GLM+HAC comparator's standard errors were if anything too conservative rather than too small. Under high persistent autocorrelation, the situation changes qualitatively: both methods' Type I error inflates substantially, but \texttt{betark}'s inflation is consistently and substantially smaller than the GLM+HAC comparator's in every comparable cell, and -- critically -- \texttt{betark}'s inflation recedes with increasing $T$ in every scenario examined, whereas the GLM+HAC comparator's inflation in the same high-persistence cells shows little improvement, and in one scenario (AR(3) high persistent) increases modestly as $T$ grows from 100 to 400. This last pattern has no straightforward asymptotic justification: the nonparametric HAC correction's bandwidth, selected according to a rule designed for generic serial dependence of unspecified form \cite{newey1994}, appears unable to keep pace with a structured, increasingly persistent, increasingly higher-order autoregressive process as efficiently as a model that estimates that exact structure directly. For \texttt{betark}, by contrast, the orderly improvement with $T$ in every scenario is consistent with ordinary asymptotic convergence of a correctly specified joint maximum likelihood estimator, with the practical caveat being that convergence is markedly slower under AR(3) high-persistent autocorrelation than under the other eight scenarios examined.

\subsection{Bias and Standard Error Ratio}

Both estimators were essentially unbiased across every condition examined, restricted to non-null effect sizes as described in Section 2.4.5. This confirms that the differences in coverage, Type I error, and power documented above arise entirely from differences in variance estimation, not from systematic error in point estimation -- the same conclusion reached for the analogous Prais--Winsten and Newey--West comparisons in continuous and panel time series settings \cite{linden2026a,linden2026b,linden2026c}. The standard error ratio results explain this pattern. \texttt{betark}'s standard errors are the ordinary asymptotic standard errors of a joint conditional maximum likelihood estimator, and they reflect the estimated autoregressive structure without any separate correction; the standard error ratio results confirm that this is, in practice, close to correctly calibrated outside the most extreme persistence conditions. The GLM+HAC comparator's standard errors, by contrast, depend on a nonparametric estimate of the long-run variance of the score that must be estimated from the same finite sample used for the point estimate; under high persistent autocorrelation this long-run variance is large relative to what a bandwidth selected by a generic rule-of-thumb can capture, producing the systematic underestimation documented here. This is consistent with the broader finding that an estimator that knows the form of the dependence it must correct for outperforms one that must estimate that form nonparametrically from the same data used for inference, precisely when the dependence is strong enough that nonparametric estimation of its long-run consequences becomes difficult.

\subsection{Misspecification and Sensitivity}

The misspecification results provide a measure of practical reassurance for \texttt{betark} users. Fitting an AR order one lag away from the true generating order -- in either direction -- carried an inferential cost of one to two percentage points in coverage, an order of magnitude smaller than the 20-plus percentage point differences associated with autocorrelation persistence and series length documented in the primary results. This robustness to moderate lag-order misspecification mirrors the analogous finding reported for Prais--Winsten-based estimators in both single-series and panel interrupted time series settings \cite{linden2026b,linden2026c}, and suggests that \texttt{betark} users need not identify the exact autoregressive order with certainty; a reasonable diagnostic-based choice, even if off by one lag, is unlikely to materially compromise inference. The sensitivity results similarly support the generality of the primary findings: \texttt{betark}'s calibration was stable across starting means ranging from a base rate as low as $5\%$ to a centrally located mean of $50\%$, and across both a flat and a sloped pre-intervention counterfactual trend, with the GLM+HAC comparator showing a consistent five-to-ten percentage point coverage deficit relative to \texttt{betark} throughout this comparatively favorable, low-persistence corner of the design space. The primary findings are therefore not an artifact of the specific starting mean or trend specification used in the main simulation grid.

\subsection{Practical Recommendations}

The results support a calibrated, condition-dependent recommendation rather than a uniform one. Under mild positive or oscillatory autocorrelation, both estimators perform reasonably well, and the choice between them is of secondary importance relative to ensuring an adequate series length. Under high persistent autocorrelation -- the condition most likely to arise in cumulative or slowly evolving health research outcomes, and the condition under which the consequences of estimator choice are most severe -- \texttt{betark} is clearly preferable to the GLM+HAC comparator at every series length examined, and the size of this advantage grows, rather than shrinks, with the order of the autoregressive process. At the same time, \texttt{betark} should not be presented as a fully solved problem under the most demanding condition examined: for AR(3) high-persistent autocorrelation specifically, even \texttt{betark}'s inference remains materially anticonservative at series lengths up to $T=400$, and researchers working with short-to-moderate proportion or rate time series under suspected high-order, high-persistence serial dependence should interpret nominal $p$-values and confidence intervals with corresponding caution regardless of which of the two estimators evaluated here is used. Researchers need not identify the exact autoregressive order before fitting \texttt{betark}; the misspecification results indicate that a defensible diagnostic-based choice, even if off by one lag, carries a substantially smaller inferential cost than the underlying persistence of the autocorrelation process itself. Regardless of estimator, researchers evaluating proportion or rate outcomes in an interrupted time series design should characterize the autocorrelation structure of their data, report the lag order used, and consider sensitivity analyses under alternative specifications, consistent with general guidance for the design and reporting of quasi-experimental program evaluations \cite{linden2005}.

\subsection{Limitations}

Several limitations of the present study should be noted. The simulation design reflects a single-group interrupted time series application with a single intervention point and a single covariate structure (intercept, time, intervention dummy, and the intervention-by-time interaction); multiple-group (controlled) designs, multiple interventions, seasonality, and covariate adjustment were not examined and represent natural extensions of this work, paralleling the multiple-group and panel extensions already developed for the analogous Prais--Winsten estimators \cite{linden2026b,linden2026c}. The starting mean was fixed at $\mu_0=.10$ in the primary design; although the sensitivity analysis confirmed that the relative ranking of the two estimators is preserved across starting means from $.05$ to $.50$, the absolute magnitude of miscalibration under high persistent autocorrelation may differ at means not examined here, particularly very close to either boundary of $(0,1)$. The maximum series length examined was $T=400$; the AR(3) high-persistent condition had not fully stabilized at this length for either estimator, and longer series may be required to fully characterize the asymptotic behavior of both methods under this condition. The maximum autoregressive order examined was $k=3$; whether \texttt{betark}'s calibration continues to degrade, and by how much, at AR(4) and beyond is unknown. The dispersion parameter was fixed via a single coefficient of variation throughout the primary design; a preliminary evaluation (Appendix~A) confirmed negligible bias in \texttt{betark}'s point estimates at this and lower dispersion levels, but the joint behavior of dispersion, AR order, and series length was not fully crossed. Finally, only two estimation approaches were evaluated: \texttt{betark} and a quasi-binomial generalized linear model with Newey–West heteroskedasticity- and autocorrelation-consistent standard errors. These represent, respectively, the only model specifically developed to jointly accommodate bounded outcomes and autocorrelated errors and the only readily available semiparametric alternative. Other approaches, such as Bayesian formulations of the beta-AR model and alternative fractional-response specifications, were beyond the scope of the present study and represent natural directions for future research.

\section{Conclusion}

This study provides the first systematic evaluation of the finite-sample inferential properties of a joint conditional maximum likelihood beta-AR($k$) model for interrupted time series analysis of proportion and rate outcomes, and the first comparison of this approach against the only other readily available estimator for this data structure: a quasi-binomial generalized linear model with Newey--West HAC standard errors. \texttt{betark} provided better-calibrated inference than the GLM+HAC comparator across the large majority of autocorrelation structures, series lengths, and effect sizes examined, with the size of this advantage widening as autoregressive order and persistence increased -- precisely the conditions under which the consequences of estimator choice are most severe in applied practice. Both methods were essentially unbiased, confirming that the differences documented here concern variance estimation and inferential calibration rather than point estimation. One important limitation was that, under the most demanding condition examined, AR(3) high-persistent autocorrelation, \texttt{betark}'s own inference remained materially anticonservative even at the largest series length evaluated, underscoring that no estimator examined here fully resolves the inferential challenge posed by strong, higher-order serial dependence in bounded outcome data. Misspecification of the autoregressive order by a single lag, and variation in the starting mean and pre-intervention trend, had only modest effects on \texttt{betark}'s performance relative to the effects of autocorrelation persistence and series length themselves. Future research should extend this evaluation to multiple-group and panel interrupted time series designs, longer series lengths and higher autoregressive orders, and alternative estimation frameworks for bounded, autocorrelated outcome data.

\bibliographystyle{unsrt}
\bibliography{betark_references}

\clearpage
\section*{Tables}

\begingroup
\small
\begin{longtable}{p{0.40\textwidth} p{0.52\textwidth}}
\caption{Simulation design and inputs.} \label{tab:design} \\
\toprule
Parameter & Input values \\
\midrule
\endfirsthead

\multicolumn{2}{c}{\tablename\ \thetable{} -- continued from previous page} \\
\toprule
Parameter & Input values \\
\midrule
\endhead

\midrule
\multicolumn{2}{r}{\emph{Continued on next page}} \\
\endfoot

\bottomrule
\endlastfoot

\multicolumn{2}{l}{\emph{Regression model}} \\
Mean equation link & Logit \\
Scale (precision) equation link & Log (constant precision; no covariates) \\
Comparator mean equation link & Logit (quasi-binomial GLM) \\
\addlinespace
\multicolumn{2}{l}{\emph{Data-generating process (equations~\ref{eq:dgplinpred}--\ref{eq:cvtophi})}} \\
Starting mean (intercept) & 0.10 \\
Dispersion & $\mathrm{cv} = .05$ ($\phi \approx 3{,}599$ at intercept $=.10$) \\
Pre-intervention trend & 0\% (flat) \\
Immediate level change & 0\% (none) \\
Post-intervention trend (treatment effect) & 0\% (null), 25\% (small), 50\% (medium), 100\% (large) \\
Intervention timing & Series midpoint \\
\addlinespace
\multicolumn{2}{l}{\emph{Series length}} \\
Number of time periods ($T$) & 100, 200, 400 \\
\addlinespace
\multicolumn{2}{l}{\emph{Autoregressive order and scenarios}} \\
AR(1) -- Scenario 1 (mild positive) & $\rho = 0.4$ \\
AR(1) -- Scenario 2 (oscillatory) & $\rho = -0.4$ \\
AR(1) -- Scenario 3 (high persistent) & $\rho = 0.7$ \\
AR(2) -- Scenario 1 (mild positive) & $\rho = (0.4, 0.2)$ \\
AR(2) -- Scenario 2 (oscillatory) & $\rho = (0.5, -0.4)$ \\
AR(2) -- Scenario 3 (high persistent) & $\rho = (0.7, 0.2)$ \\
AR(3) -- Scenario 1 (mild positive) & $\rho = (0.4, 0.2, 0.1)$ \\
AR(3) -- Scenario 2 (oscillatory) & $\rho = (0.7, -0.3, 0.15)$ \\
AR(3) -- Scenario 3 (high persistent) & $\rho = (0.6, 0.25, 0.1)$ \\
\addlinespace
\multicolumn{2}{l}{\emph{Estimators compared}} \\
\texttt{betark} & Joint conditional ML, beta-AR($k$); \texttt{lag()} matched to true AR order \\
\texttt{glm} + HAC & Quasi-binomial GLM, logit link; \texttt{vce(hac nwest \#)}, \texttt{\#} matched to true AR order \\
\addlinespace
\multicolumn{2}{l}{\emph{Misspecification analysis}} \\
Design & DGP = true AR order; \texttt{betark} fitted with one lag under-specified and one lag over-specified \\
Autocorrelation scenario & Mild positive only (Scenario 1, each AR order) \\
$T$ & 100, 200, 400 \\
Post-intervention trend & 50\% \\
\addlinespace
\multicolumn{2}{l}{\emph{Sensitivity analysis}} \\
Design & AR(1) Scenario 1 ($\rho=0.4$), $T=200$, held fixed \\
Starting mean (intercept) & 0.05, 0.10, 0.30, 0.50 \\
Pre-intervention trend & $-25\%$, 0\%, $25\%$ \\
Post-intervention trend & 0\% (null), 50\% (medium) \\
\addlinespace
\multicolumn{2}{l}{\emph{Performance measures} \cite{burton2006}} \\
Statistical power & Proportion of replications rejecting $H_0: \beta_{\mathrm{post}}-\beta_{\mathrm{pre}}=0$ at $\alpha=.05$ when the true effect is non-zero (evaluated separately at each non-null effect size) \\
95\% CI coverage & Proportion of replications in which the 95\% CI contains the true effect; non-null effect sizes only, averaged \\
Type I error rate & Proportion of replications rejecting $H_0$ when the true effect is zero; null condition only \\
Percentage bias & Mean(estimate) $-$ true value; non-null effect sizes only, averaged \\
Standard error (SE) ratio & Mean(model-based SE) / SD(point estimates); non-null effect sizes only, averaged; values $<1$ indicate anticonservative inference \\
\addlinespace
\multicolumn{2}{l}{\emph{Monte Carlo settings}} \\
Replications per condition & 2000 \\
Significance level ($\alpha$) & 0.05 (two-sided Wald test) \\
Total design conditions & $3 \text{(}T\text{)} \times 9 \text{(scenarios)} \times 4 \text{(effect sizes)} = 108$ \\

\end{longtable}
\endgroup

\clearpage
\begin{table}[htbp]
\centering
\caption{Statistical power, small effect size (25\% trend). Proportion of replications rejecting $H_0$ at $\alpha=.05$.}
\label{tab:power25}
\resizebox{\textwidth}{!}{%
\begin{tabular}{lccc ccc}
\toprule
 & \multicolumn{3}{c}{\texttt{betark}} & \multicolumn{3}{c}{\texttt{glm} + HAC(nwest)} \\
\cmidrule(lr){2-4} \cmidrule(lr){5-7}
Scenario & $T=100$ & $T=200$ & $T=400$ & $T=100$ & $T=200$ & $T=400$ \\
\midrule
AR(1) mild positive & 0.992 & 1.000 & 1.000 & 1.000 & 1.000 & 1.000 \\
AR(1) oscillatory & 1.000 & 1.000 & 1.000 & 1.000 & 1.000 & 1.000 \\
AR(1) high persistent & 0.708 & 0.834 & 0.986 & 0.914 & 0.962 & 0.996 \\
\addlinespace
AR(2) mild positive & 0.846 & 0.976 & 0.998 & 0.972 & 0.996 & 1.000 \\
AR(2) oscillatory & 1.000 & 1.000 & 1.000 & 1.000 & 1.000 & 1.000 \\
AR(2) high persistent & 0.466 & 0.386 & 0.456 & 0.720 & 0.758 & 0.858 \\
\addlinespace
AR(3) mild positive & 0.746 & 0.888 & 0.976 & 0.910 & 0.972 & 0.996 \\
AR(3) oscillatory & 0.912 & 0.998 & 1.000 & 0.978 & 1.000 & 1.000 \\
AR(3) high persistent & 0.542 & 0.400 & 0.336 & 0.738 & 0.738 & 0.788 \\
\bottomrule
\end{tabular}%
}
\end{table}

\clearpage
\begin{table}[htbp]
\centering
\caption{Statistical power, medium effect size (50\% trend).}
\label{tab:power50}
\resizebox{\textwidth}{!}{%
\begin{tabular}{lccc ccc}
\toprule
 & \multicolumn{3}{c}{\texttt{betark}} & \multicolumn{3}{c}{\texttt{glm} + HAC(nwest)} \\
\cmidrule(lr){2-4} \cmidrule(lr){5-7}
Scenario & $T=100$ & $T=200$ & $T=400$ & $T=100$ & $T=200$ & $T=400$ \\
\midrule
AR(1) mild positive & 1.000 & 1.000 & 1.000 & 1.000 & 1.000 & 1.000 \\
AR(1) oscillatory & 1.000 & 1.000 & 1.000 & 1.000 & 1.000 & 1.000 \\
AR(1) high persistent & 0.978 & 1.000 & 1.000 & 0.998 & 1.000 & 1.000 \\
\addlinespace
AR(2) mild positive & 1.000 & 1.000 & 1.000 & 1.000 & 1.000 & 1.000 \\
AR(2) oscillatory & 1.000 & 1.000 & 1.000 & 1.000 & 1.000 & 1.000 \\
AR(2) high persistent & 0.684 & 0.686 & 0.804 & 0.916 & 0.934 & 0.978 \\
\addlinespace
AR(3) mild positive & 0.984 & 1.000 & 1.000 & 1.000 & 1.000 & 1.000 \\
AR(3) oscillatory & 1.000 & 1.000 & 1.000 & 1.000 & 1.000 & 1.000 \\
AR(3) high persistent & 0.662 & 0.593 & 0.578 & 0.862 & 0.872 & 0.904 \\
\bottomrule
\end{tabular}%
}
\end{table}

\clearpage
\begin{table}[htbp]
\centering
\caption{Statistical power, large effect size (100\% trend).}
\label{tab:power100}
\resizebox{\textwidth}{!}{%
\begin{tabular}{lccc ccc}
\toprule
 & \multicolumn{3}{c}{\texttt{betark}} & \multicolumn{3}{c}{\texttt{glm} + HAC(nwest)} \\
\cmidrule(lr){2-4} \cmidrule(lr){5-7}
Scenario & $T=100$ & $T=200$ & $T=400$ & $T=100$ & $T=200$ & $T=400$ \\
\midrule
AR(1) mild positive & 1.000 & 1.000 & 1.000 & 1.000 & 1.000 & 1.000 \\
AR(1) oscillatory & 1.000 & 1.000 & 1.000 & 1.000 & 1.000 & 1.000 \\
AR(1) high persistent & 1.000 & 1.000 & 1.000 & 1.000 & 1.000 & 1.000 \\
\addlinespace
AR(2) mild positive & 1.000 & 1.000 & 1.000 & 1.000 & 1.000 & 1.000 \\
AR(2) oscillatory & 1.000 & 1.000 & 1.000 & 1.000 & 1.000 & 1.000 \\
AR(2) high persistent & 0.914 & 0.948 & 0.990 & 1.000 & 1.000 & 1.000 \\
\addlinespace
AR(3) mild positive & 1.000 & 1.000 & 1.000 & 1.000 & 1.000 & 1.000 \\
AR(3) oscillatory & 1.000 & 1.000 & 1.000 & 1.000 & 1.000 & 1.000 \\
AR(3) high persistent & 0.888 & 0.812 & 0.794 & 0.980 & 0.984 & 0.986 \\
\bottomrule
\end{tabular}%
}
\end{table}

\clearpage
\begin{table}[htbp]
\centering
\caption{95\% confidence interval coverage, averaged across non-null trend effect sizes (25\%, 50\%, 100\%). Null-effect replications are reserved for Type I error (Table~\ref{tab:typeI}) and excluded here. Nominal value is 0.95.}
\label{tab:coverage}
\resizebox{\textwidth}{!}{%
\begin{tabular}{lccc ccc}
\toprule
 & \multicolumn{3}{c}{\texttt{betark}} & \multicolumn{3}{c}{\texttt{glm} + HAC(nwest)} \\
\cmidrule(lr){2-4} \cmidrule(lr){5-7}
Scenario & $T=100$ & $T=200$ & $T=400$ & $T=100$ & $T=200$ & $T=400$ \\
\midrule
AR(1) mild positive & 0.9080 & 0.9240 & 0.9440 & 0.8347 & 0.8453 & 0.8547 \\
AR(1) oscillatory & 0.9480 & 0.9500 & 0.9460 & 0.9753 & 0.9820 & 0.9787 \\
AR(1) high persistent & 0.8533 & 0.9093 & 0.9340 & 0.6387 & 0.6727 & 0.7113 \\
\addlinespace
AR(2) mild positive & 0.8600 & 0.9093 & 0.9273 & 0.7233 & 0.7667 & 0.7780 \\
AR(2) oscillatory & 0.9253 & 0.9407 & 0.9520 & 0.9667 & 0.9773 & 0.9840 \\
AR(2) high persistent & 0.6773 & 0.7780 & 0.8607 & 0.4287 & 0.4573 & 0.4627 \\
\addlinespace
AR(3) mild positive & 0.7907 & 0.8527 & 0.9220 & 0.6673 & 0.6753 & 0.7387 \\
AR(3) oscillatory & 0.8633 & 0.9173 & 0.9287 & 0.8227 & 0.8520 & 0.8733 \\
AR(3) high persistent & 0.5653 & 0.6771 & 0.7353 & 0.3447 & 0.3313 & 0.3420 \\
\bottomrule
\end{tabular}%
}
\end{table}

\clearpage
\begin{table}[htbp]
\centering
\caption{Type I error rate, evaluated under the true null (0\% trend effect). Nominal value is 0.05.}
\label{tab:typeI}
\resizebox{\textwidth}{!}{%
\begin{tabular}{lccc ccc}
\toprule
 & \multicolumn{3}{c}{\texttt{betark}} & \multicolumn{3}{c}{\texttt{glm} + HAC(nwest)} \\
\cmidrule(lr){2-4} \cmidrule(lr){5-7}
Scenario & $T=100$ & $T=200$ & $T=400$ & $T=100$ & $T=200$ & $T=400$ \\
\midrule
AR(1) mild positive & 0.0840 & 0.0760 & 0.0540 & 0.1740 & 0.1660 & 0.1400 \\
AR(1) oscillatory & 0.0660 & 0.0660 & 0.0580 & 0.0380 & 0.0300 & 0.0220 \\
AR(1) high persistent & 0.1320 & 0.0960 & 0.0720 & 0.3520 & 0.3360 & 0.2680 \\
\addlinespace
AR(2) mild positive & 0.1420 & 0.1180 & 0.0760 & 0.2660 & 0.2740 & 0.2160 \\
AR(2) oscillatory & 0.0580 & 0.0660 & 0.0540 & 0.0180 & 0.0280 & 0.0100 \\
AR(2) high persistent & 0.3080 & 0.1820 & 0.1480 & 0.6020 & 0.5240 & 0.5380 \\
\addlinespace
AR(3) mild positive & 0.2200 & 0.1380 & 0.0960 & 0.3520 & 0.3100 & 0.2860 \\
AR(3) oscillatory & 0.1380 & 0.1060 & 0.0600 & 0.1880 & 0.1700 & 0.1220 \\
AR(3) high persistent & 0.4300 & 0.3400 & 0.2300 & 0.6540 & 0.6700 & 0.6260 \\
\bottomrule
\end{tabular}%
}
\end{table}

\clearpage
\begin{table}[htbp]
\centering
\caption{Bias (mean estimate minus true value), averaged across non-null trend effect sizes (25\%, 50\%, 100\%). Null-effect replications are excluded; see Table~\ref{tab:typeI} for null-condition results.}
\label{tab:bias}
\resizebox{\textwidth}{!}{%
\begin{tabular}{lccc ccc}
\toprule
 & \multicolumn{3}{c}{\texttt{betark}} & \multicolumn{3}{c}{\texttt{glm} + HAC(nwest)} \\
\cmidrule(lr){2-4} \cmidrule(lr){5-7}
Scenario & $T=100$ & $T=200$ & $T=400$ & $T=100$ & $T=200$ & $T=400$ \\
\midrule
AR(1) mild positive & 0.0001 & -0.0000 & 0.0000 & 0.0001 & -0.0000 & 0.0000 \\
AR(1) oscillatory & 0.0000 & 0.0000 & 0.0000 & 0.0000 & 0.0000 & -0.0000 \\
AR(1) high persistent & 0.0002 & -0.0000 & 0.0000 & 0.0002 & 0.0000 & 0.0000 \\
\addlinespace
AR(2) mild positive & -0.0000 & 0.0000 & -0.0000 & -0.0000 & 0.0000 & -0.0000 \\
AR(2) oscillatory & 0.0000 & 0.0000 & 0.0000 & 0.0000 & -0.0000 & -0.0000 \\
AR(2) high persistent & 0.0001 & 0.0000 & -0.0000 & 0.0001 & 0.0001 & 0.0001 \\
\addlinespace
AR(3) mild positive & -0.0001 & 0.0000 & -0.0000 & -0.0001 & 0.0000 & -0.0000 \\
AR(3) oscillatory & 0.0000 & 0.0000 & -0.0000 & 0.0000 & 0.0000 & -0.0000 \\
AR(3) high persistent & 0.0004 & 0.0002 & 0.0001 & 0.0006 & 0.0003 & 0.0001 \\
\bottomrule
\end{tabular}%
}
\end{table}

\clearpage
\begin{table}[htbp]
\centering
\caption{Standard error ratio (mean model-based SE / SD of point estimates), averaged across non-null trend effect sizes (25\%, 50\%, 100\%). Null-effect replications are excluded; see Table~\ref{tab:typeI} for null-condition results.}
\label{tab:seratio}
\resizebox{\textwidth}{!}{%
\begin{tabular}{lccc ccc}
\toprule
 & \multicolumn{3}{c}{\texttt{betark}} & \multicolumn{3}{c}{\texttt{glm} + HAC(nwest)} \\
\cmidrule(lr){2-4} \cmidrule(lr){5-7}
Scenario & $T=100$ & $T=200$ & $T=400$ & $T=100$ & $T=200$ & $T=400$ \\
\midrule
AR(1) mild positive & 0.914 & 0.955 & 0.987 & 0.724 & 0.744 & 0.768 \\
AR(1) oscillatory & 0.977 & 1.008 & 1.017 & 1.141 & 1.186 & 1.195 \\
AR(1) high persistent & 0.791 & 0.880 & 0.944 & 0.472 & 0.508 & 0.527 \\
\addlinespace
AR(2) mild positive & 0.801 & 0.898 & 0.930 & 0.575 & 0.613 & 0.625 \\
AR(2) oscillatory & 0.927 & 0.987 & 1.009 & 1.113 & 1.183 & 1.217 \\
AR(2) high persistent & 0.573 & 0.818 & 0.662 & 0.285 & 0.301 & 0.306 \\
\addlinespace
AR(3) mild positive & 0.675 & 0.787 & 0.920 & 0.490 & 0.514 & 0.570 \\
AR(3) oscillatory & 0.821 & 0.919 & 0.959 & 0.706 & 0.766 & 0.787 \\
AR(3) high persistent & 0.547 & 0.771 & 0.774 & 0.233 & 0.227 & 0.220 \\
\bottomrule
\end{tabular}%
}
\end{table}

\clearpage
\begin{table}[htbp]
\centering
\caption{Illustrative example: slope-change coefficient (logit scale) by AR order and method.}
\label{tab:applied}
\begin{tabular}{llccc c}
\toprule
AR order & Method & Coefficient & SE & $p$-value & 95\% CI \\
\midrule
AR(1) & \texttt{betark} & $-0.00266$ & $0.00059$ & $<0.001$ & $(-0.00380,\ -0.00151)$ \\
      & \texttt{glm}+HAC(1) & $-0.00263$ & $0.00037$ & $<0.001$ & $(-0.00336,\ -0.00191)$ \\
\addlinespace
AR(2) & \texttt{betark} & $-0.00276$ & $0.00134$ & $0.039$ & $(-0.00537,\ -0.00014)$ \\
      & \texttt{glm}+HAC(2) & $-0.00222$ & $0.00060$ & $<0.001$ & $(-0.00340,\ -0.00105)$ \\
\addlinespace
AR(3) & \texttt{betark} & $-0.00182$ & $0.00170$ & $0.283$ & $(-0.00516,\ 0.00151)$ \\
      & \texttt{glm}+HAC(3) & $-0.00103$ & $0.00069$ & $0.134$ & $(-0.00238,\ 0.00032)$ \\
\bottomrule
\end{tabular}
\end{table}

\clearpage
\section*{Figures}

\end{document}